\documentclass[dvips]{acta}
\usepackage{supertabular,lscape,epsfig}
\usepackage{amssymb}
\usepackage{amsmath}
\SetPages{1}{6}
\SetVol{73}{2023}

\begin{document}

\begin{Titlepage}

\Title { AM CVn -- System Parameters and Gravitational Waves }

\Author {J.~~S m a k}
{N. Copernicus Astronomical Center, Polish Academy of Sciences,\\
Bartycka 18, 00-716 Warsaw, Poland\\
e-mail: jis@camk.edu.pl }

\Received{  }

\end{Titlepage}

\Abstract { System parameters are re-determined: $M_1=0.86\pm0.18M\odot$,  
$M_2=0.103\pm0.022M\odot$, $A=1.508\pm 0.100\times 10^{10}$cm, 
and $i=69\pm3^{\circ}$. 
The secondary component is a semi-degenerate helium star loosing mass 
at a rate $\dot M=4.93\pm 1.65\times10^{-9}M\odot/yr$. 
The accretion disk is sufficiently hot to avoid thermal instability.  

The orbital light curve recovered from observations made in 1962 
shows minimum shifted to phase $\phi=0.50$, corresponding to $O-C=0.0060$d. 
Together with mimima observed in 1992-99 this implies that the orbital 
period is increasing at a rate $dP/dt\approx 8.5 \times 10^{-13}$ 
consistent with predictions involving the emission of gravitational waves. 
 } 

{binaries: cataclysmic variables, stars: individual: AM CVn, gravitational waves }

%Sec.1
\section { Introduction } 

AM CVn (earlier known as HZ 29; Humason and Zwicky 1947) is a prototype 
of the ultra short period, helium cataclysmic variables (Nelemans 2005, 
Solheim 2010, Kotko {\it et al.} 2012). 
Its light variations were discovered in 1962 (Smak 1967). Their period  
of only $P\approx 18$ minutes was originally interpreted as the orbital period 
and Paczy{\'n}ski (1967), in his pioneering paper on graviational waves 
and the evolution of close binaries, predicted that {\it "this system could be 
most easily used to check the existence of gravitational radiation"}. 
Regretfully, the light variations of AM CVn turned out to be quite complicated. 
So complicated that it took many decades and several, extensive photometric   
and spectroscopic studies (Patterson {\it et al.} 1993, Harvey {\it et al.} 1998, 
Skillman {\it et al.} 1999, Nelemans {\it et al.} 2001a, Roelofs {\it et al.} 
2006 and references therein) before they were fully interpreted. 
Thanks to those investigations the light variations of AM CVn are now known 
to be a superposition of the three components: 
the superhumps with $P_{SH}=1051.2s$, the negative superhumps with 
$P_{nSH}=1011.4s$ (both of them with variable amplitude), and variations 
with the orbital period $P_{orb}=1028.7s$. 
The superhumps are the dominant component and their period is the main 
observed period. 
It may be added that light variations of AM CVn, when discovered in 1962, 
were the first -- unrecognized at that time(!) -- example of superhumps. 

System parameters of AM CVn were determined by Roelofs {\it et al.} (2006). 
As the two main constraints they used (1) the amplitude of the radial velocity 
curve of the primary component $K_1=92\pm4$ km/s and (2) the mass ratio $q=0.18$, 
determined from $K_1$ and the amplitude $A_S=471\pm11$ km/s of the S-wave 
observed in many lines in the spectrum of AM CVn, by assuming that it represents 
the velocity of the stream at the point where the hot spot is formed. 
The orbital inclination $i=43\pm2^{\circ}$ was obtained by requiring that the predicted absolute magnitude be equal to the absolute magnitude of AM CVn. Crucial for this 
step was the use of the HST parallax ($\pi=1.65\pm0.30$ mas; Roelofs {\it et al.} 2007a). 
Shortly afterwards, however, when the much larger Gaia parallax 
($\pi=3.351\pm0.045$ mas) became available (Ramsay {\it et al.} 2018), it was 
obvious that those elements must be significantly revised. 

The aim of the present paper is twofold: (1) to re-determine the system 
parameters (Section 2) and (2) to perform the belated(!) test for 
the existence of gravitational waves using the evidence provided by 
the orbital light curve recovered from the earliest 1962 photometry (Section 3).

%Sec.2
\section { System Parameters } 

%2.1
\subsection { The Mass Ratio }

Roelofs {\it et al.} (2006), neglecting the evidence from the superhump and negative 
superhump periods, determined the mass ratio from $K_1$ and the amplitude 
of the S-wave $A_S$ by assuming that it represents the velocity of the stream 
at the point where the hot spot is formed. This assumption, however, is far from 
being safe. Worth recalling are two well documented cases: WZ Sge and U Gem.  
The amplitude of the S-wave observed in the spectrum of WZ Sge (Krzemi{\'n}ski 
and Kraft 1964) is variable between $A_S=650$ and 850 km/s. In the case of
U Gem (Smak 1976) the parameters of the S-wave determined from different 
spectral lines are different, for example:   
$A_S=340$ km/s and $\phi_o=0.126$ from the CaII K line {\it versus} 
$A_S=404$ km/s and $\phi_o=0.069$ from the HeI $\lambda$4471 line. 

We determine the mass ratio from the superhump period excess and the negative 
superhump period deficiency. This requires an extra comment. 
There has been a debate whether the well defined and well understood 
$q-\epsilon_{SH}$ relation existing for the hydrogen rich cataclysmic variables
can be applied to the AM CVn systems (e.g. Pearson 2007). Fortunately we have 
the evidence from YZ LMi = SDSS J0926+3624 (Copperwheat {\it et al.} 2011) 
showing that the mass ratios obtained from superhumps and from the analysis 
of eclipses are practically identical. And, in addition to superhumps, we have 
the negative superhumps for which a similar $q-\epsilon_{nSH}$ relation is 
well established. 

Using the values of the three periods: $P_{SH}$, $P_{nSH}$ and $P_{orb}$ 
(Skillman {\it et al.} 1999) we obtain $\epsilon_{SH}=0.0218$ and 
$\epsilon_{nSH}=0.0168$. 
The $q-\epsilon_{SH}$ relation from Kato {\it et al.} (2009, Fig.31) 

%Eq.1
\beq
\epsilon_{SH}~=~0.16~q~+~0.25~q^2
\eeq

\noindent
gives: $q=0.116$. For the negative superhumps we have the theoretical relation 
from Osaki and Kato (2013, Eq.1) 

%Eq.2
\beq
{\epsilon_{nSH}\over {1-\epsilon_{nSH}}}~=~{3\over 7}~{q\over {(1+q)^{1/2}}}~r_d^{3/2}
\eeq

\noindent
which gives $q=0.122$. The agreement between the two determinations is near perfect 
and we adopt $q=0.120\pm 0.005$.

%2.2
\subsection { The Orbital Inclination }

To begin with we may ask what can be learned about the inclination 
from the orbital light curve (Skillman {\it et al.} 1999, Fig.6). Its shape 
is most likely due to the irradiation of the secondary component, but its 
amplitude ($\sim 0.01$mag.) is too small to provide any meaningful 
information on inclination. 

We follow the approach taken by Roelofs {\it et al.} (2006) and determine 
the orbital inclination by using, as an additional constraint, the observed  
magnitude of AM CVn. Following the procedures described in Sections 2.3 and 
2.4 we calculate system parameters as functions of inclination and then 
require that the {\it predicted} V magnitude (using the Gaia parallax: 
$\pi=3.351\pm0.045$ mas; Ramsay {\it et al.} 2018) be equal to the 
{\it true} magnitude: $<V>\approx 14.1$ (cf. Patterson {\it et al.} 1992).  
Results, shown in Fig.1, give $i=69\pm3^{\circ}$.

%***Fig.1
\begin{figure}[htb]
\includegraphics{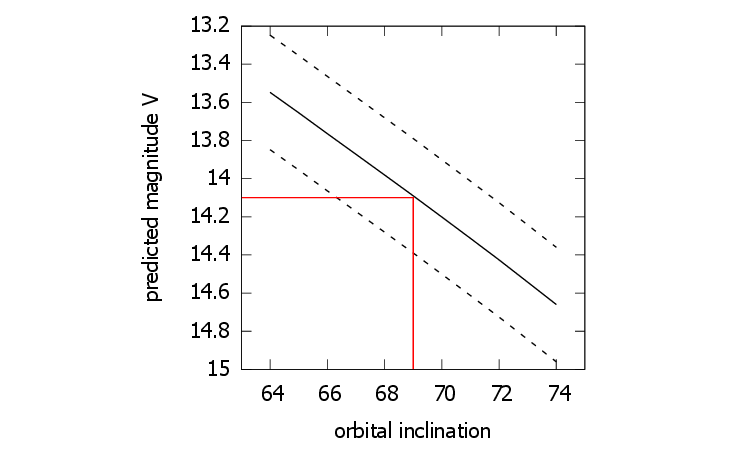}
\FigCap { The $V-i$ relation. Dotted lines show formal errors. The red lines
show how $i=69\pm 3^{\circ}$ is determined. See text for details. }
\end{figure}

%2.3
\subsection { System Parameters }

Using $K_1=92\pm4$ km/s, $q=0.120\pm 0.005$ and $i=69\pm3^{\circ}$ we obtain:  
$A_{orb}=1.508\pm 0.100\times 10^{10}$cm, $M_1=0.86\pm0.18M\odot$ and 
$M_2=0.103\pm0.022M\odot$.
Furthermore we also get the radius of the secondary component
$R_2=0.047\pm0.003R\odot$ and note that it satisfies the $M-R$ relation 
for semi-degenerate secondaries (Nelemans {\it et al.} 2001b, 
Roelofs {\it et al.} 2007b) 

%Eq.3
\beq
R/R\odot~=~0.043~M/M\odot^{-0.062}~.
\eeq

%2.4
\subsection { The Mass Transfer Rate and the Absolute Magnitude }

The emission of gravitational waves results in the loss of angular momentum,  
the loss of mass from the secondary component and its transfer to the primary 
(cf. Paczy{\'n}ski 1967, Roelofs {\it et al.} 2006). The relevant formulae and results are: 

%Eq.4
\beq
{{d\ln J}\over {dt}}~=~-{{32}\over{5}} {{G^{5/3}}\over {c^5}}
\left( {{P}\over {2\pi}}\right )^{-8/3} {{M_1M_2}\over {(M_1+M_2)^{1/3}}}~
=~-~1.015\pm 0.258 \times 10^{-15}~, 
\eeq

%Eq.5
\beq
{{d\ln M_2}\over {dt}}~=~{{2}\over {\zeta_2+5/3-2q}} {{d\ln J}\over {dt}}~
=~-~1.488\pm 0.378 \times 10^{-15}~,
\eeq

\noindent 
where $\zeta_2=-0.062$ is the exponent in the $M-R$ relation for semi-degenerate 
secondaries (cf. Eq.3). The corresponding mass transfer rate is then 
${\dot M}=4.93\pm 1.65\times10^{-9}M\odot/yr$. 

The accretion disk is quite hot. Using standard equation 

%*Eq.6
\beq
\sigma T_e^4~=~{3\over{8\pi}}~{{GM_1}\over {R^3}}~\dot M~
		 \left [~1~-~\left({{R_1}\over {R}}\right)^{1/2}~\right ]~,
\eeq 

\noindent
we find that at the outer edge of the disk (i.e. at $r_d=0.88r_{Roche}=0.50$) 
the temperature is $\log T_e=4.27$. This is safely higher than 
$\log T_{e,crit}=4.1$ below which the helium accretion disk becomes thermally 
unstable (cf. Smak 1983). 

The accretion disk is also quite bright. To calculate its absolute magnitude 
we integrate fluxes over its two sides with local temperatures being determined 
from Eq.6 and correct the resulting luminosity for inclination using 

%Eq.7
\beq
L_d(i)~=~<L_d> {6\over {3-u}} (1-u+u\cos i)\cos i~, 
\eeq

\noindent
with limb darkening coefficient $u=0.6$ (note that this relation is valid for 
$i\ll 90^{\circ}$). 

The results (see also Subsection 2.2) are: $M_V=6.72\pm 0.30$ and (with 
$\pi=3.351\pm0.045$ mas) $V=14.09\pm 0.30$, which agrees with the {\it true} 
magnitude: $<V>\approx 14.1$ (cf. Patterson {\it et al.} 1992).

%2.5
\subsection { The Orbital Period }

The orbital peiod is increasing (cf. Paczy{\'n}ski 1967) at a rate 

%Eq.8
\beq
{{d\ln P}\over {dt}}~=~3~\left[1~-~{{2(1-q)}\over{\zeta_2+5/3-2q}}\right ]~
{{d\ln J}\over {dt}}~=~8.82\pm 2.24\times 10^{-16}~, 
\eeq

\noindent
or 

%Eq.9
\beq
{{dP}\over {dt}}~=~9.14\pm 2.30\times 10^{-13}~.
\eeq

%Sec.3
\section { The 1962 Orbital Light Curve } 

We recover the mean orbital light curve from the oldest set of photometric 
observations of AM CVn made in 1962 (Smak 1967, Tables 2a and 2b) by calculating  
their phases from the elements given by Skillman {\it et al.} (1999, Eq.1) 

%Eq.10
\beq
Min~=~HJD~2448742.5610(4)~~+~0.011906623(3)~E~. 
\eeq

\noindent
and forming normal points.  

The resulting light curve is shown in Fig.2. In spite of large scatter 
its shape is satisfactorily similar to the light curves obtained by 
Skillman {\it et al.} (1999, Fig.6). The only {\it significant} difference is 
the position of the minimum at phase $\phi_{min}\approx 0.50$, 
corresponding to $(O-C)\approx 0.0060$d. 

%***Fig.2
\begin{figure}[htb]
\includegraphics{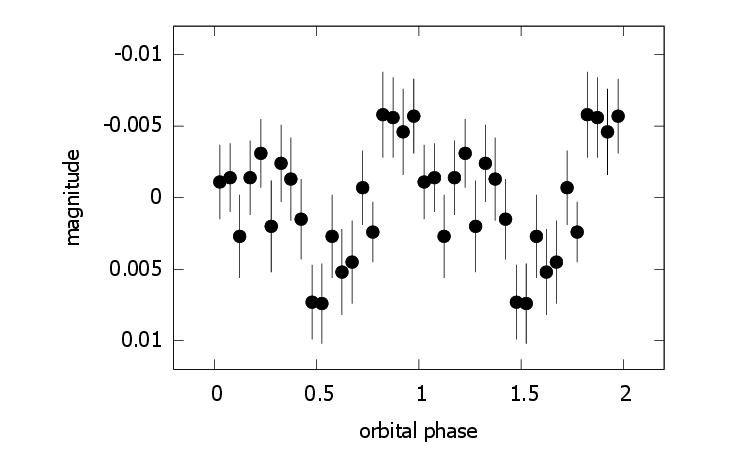}
\FigCap { The mean orbital light curve of AM CVn observed in 1962. }
\end{figure}

Adding the 1962 minimum to the minima observed by Skillman {\it et al.} 
(1999, Table 3) we determine the new light elements

%Eq.11
\beq
Min~=~HJD~2448742.5610(4)~~+~0.011906621(2)~E~+~A~E^2~, 
\eeq

\noindent
with 

%Eq.12
\beq
A~=~5.06\pm2.84\times 10^{-15}~, 
\eeq

\noindent
giving

%Eq.13
\beq
{{dP}\over {dt}}~=~8.50\pm4.77\times 10^{-13}~.
\eeq

The observed value of $dP/dt$ agrees -- within errors -- with its predicted 
value (Eq.9). 
This could be treated as another test on the existence of gravitational waves. 
Regretfully, such a test is hardly needed today. 

\begin {references} 

% \ApJ  \PASP  \MNRAS  \Acta  \AA 

\refitem {Copperwheat, C.M., Marsh, T.R., Littlefair, S.P., Dhillon, V.S., 
          Ramsay, G., Drake, A.J., G{\"a}nsicke, B.T., Groot, P.J., Hakala, P., 
          Koester, D., Nelemans, G., Roelofs, G., Southworth, J., Steeghs, D., 
          Tulloch, S.} {2011} {\MNRAS} {410} {1113}
%         q(SH) is OK

\refitem {Harvey, D.A., Skillman, D.R., Kemp, J., Patterson, J., Vanmunster, T., 
          Fried, R.E., Retter, A.} {1998} {\ApJ} {493} {L105}
%         Porb

\refitem {Humason, M., Zwicky, F.} {1947} {\ApJ} {105} {85}

\refitem {Kato, T. {\it et al.}} {2009} {\it Publ.Astr.Soc.Japan} {61} {S395} 
%         eps-q

\refitem {Kotko, I., Lasota, J.-P., Dubus, G., Hameury, J.-M.} {2012} 
         {\AA} {544} {A13}

\refitem {Krzemi{\'n}ski, W., Kraft, R.P.} {1964} {\ApJ} {140} {921}

\refitem {Nelemans, G.} {2005} {\it ASP Conference Series} {330} {27}
%        review

\refitem {Nelemans, G., Steeghs, D., Groot, P.J.} {2001a} {\AA} {326} {621}
%        spectr. Porb

\refitem {Nelemans, G., Portegies Zwart, S.F., Verbunt, F., Yungelson, L.R.} 
          {2001b} {\AA} {368} {939}
%        formulae

\refitem {Osaki, Y., Kato, T.} {2013} {\it Publ.Astr.Soc.Japan} {65} {95}
%         eps-q for nSH

\refitem {Paczy{\'n}ski, B.} {1967} {\Acta} {17} {287} 

\refitem {Patterson, J., Sterner, E., Halpern, J.P., Raymond, J.C.}
         {1992} {\ApJ} {384} {234} 
%        <V>

\refitem {Patterson, J., Halpern, J., Shambrook, A.} {1993} {\ApJ} {419} {803}
%        first evidence for Porb

\refitem {Pearson, K.J.} {2007} {\MNRAS} {379} {183}
%        eps-q ???

\refitem {Ramsay, G., Green, M.J., Marsh, T.R., Kupfer, T., Breedt, E., Korol, V.,
          Groot, P.J., Knigge, C., Nelemans, G., Steeghs, D., Woudt, P., 
          Aungwerojwit, A.} {2018} {\AA} {620} {141}
%         Gaia parallax

\refitem {Roelofs, G.H.A., Groot, P.J., Nelemans, G., Marsh, T.R., Steeghs, D.}
         {2006} {\MNRAS} {371} {1231}
%         elements

\refitem {Roelofs, G.H.A., Groot, P.J., Benedict, G.F., McArthur, B.E., 
          Steeghs, D.,Morales-Rueda, L., Marsh, T.R., Nelemans, G.}
         {2007a} {\ApJ} {666} {1174}
%        HST parallax

\refitem {Roelofs, G.H.A., Groot, P.J., Benedict, G.F., McArthur, B.E., 
          Steeghs, D.,Morales-Rueda, L., Marsh, T.R., Nelemans, G.}
         {2007b} {\it ASP Conference Series} {372} {437}
%         nature of secondary

\refitem {Skillman, D.R., Patterson, J., Kemp, J., Harvey, D.A., Fried, R.E.,
          Retter, A., Lipkin, Y., Vanmunster, T.} {1999} {\PASP} {111} {1281}
%        3xPeriods

\refitem {Smak, J.} {1967} {\Acta} {17} {256} 

\refitem {Smak, J.} {1976} {\Acta} {26} {277} 
%        UGem

\refitem {Smak, J.} {1983} {\Acta} {33} {333} 
%        log Te,crit = 4.1 

\refitem {Solheim, J.-E.} {2010} {\PASP} {122} {1133}
%        rev

\end {references}

\end{document}